\documentclass[useAMS,usegraphicx,usenatbib]{mn2e}

\title[Mapping the three-body system]{Mapping the three-body system -- decay time and reversibility}

\author[H.J. Lehto et al.]{H. J. Lehto$^{1}$\thanks{E-mail:hlehto@utu.fi}, S. Kotiranta$^{1}$, 
M. J. Valtonen$^{1}$, P. Hein\"am\"aki$^{1}$, S. Mikkola$^{1}$, A. D. Chernin$^{1,2}$\\
$^{1}$Department of physics and Tuorla observatory, University of Turku, Finland\\
$^{2}$Sternberg Astronomical Institute, Moscow university, 119899, Moscow Russia}

\begin{document}

\date{Accepted {\ldots} . Received {\ldots} ; in original form 2008
February 28 }

\pagerange{\pageref{firstpage}--\pageref{lastpage}} \pubyear{2008}

\maketitle

\begin{abstract}
In this paper we carry out a quantitative analysis of the three-body
systems and map them as a function of decaying time and intial conguration, look at
this problem as an example of a simple deterministic system, and ask to what extent the
orbits are really predictable. We have investigated the behavior of about 200 000
general Newtonian three body systems using the simplest initial conditions. Within our
resolution these cover all the possible states where the objects are initially at rest
and have no angular momentum. We have determined the decay time-scales of the triple
systems and show that the distribution of this parameter is fractal in appearance. Some
areas that appear stable on large scales exhibit very narrow strips of instability and
the overall pattern, dominated by resonances, reminds us of a traditional Maasai warrior
shield. Also an attempt is made to recover the original starting conguration of the
three bodies by backward integration. We find there are instances where the evolution
to the future and to the past lead to different orbits, in spite of
time symmetric initial conditions.  This implies that even in simple
deterministic systems there exists an Arrow of Time.

\end{abstract}


\section{Introduction} 
Dynamical interaction of multiple objects is common in physics and astronomy.
It occurs over a huge range of scales from atoms in molecules to galaxies. 
If we know the interaction potential, then it is possible in principle
to calculate the evolution of such a system. As is well known, systems
of this kind tend to be chaotic, and                    
so is the solution of the three-body problem \citep{pri,aar}.

\begin{figure*}
   \includegraphics[width=1.8\columnwidth, angle=-90]{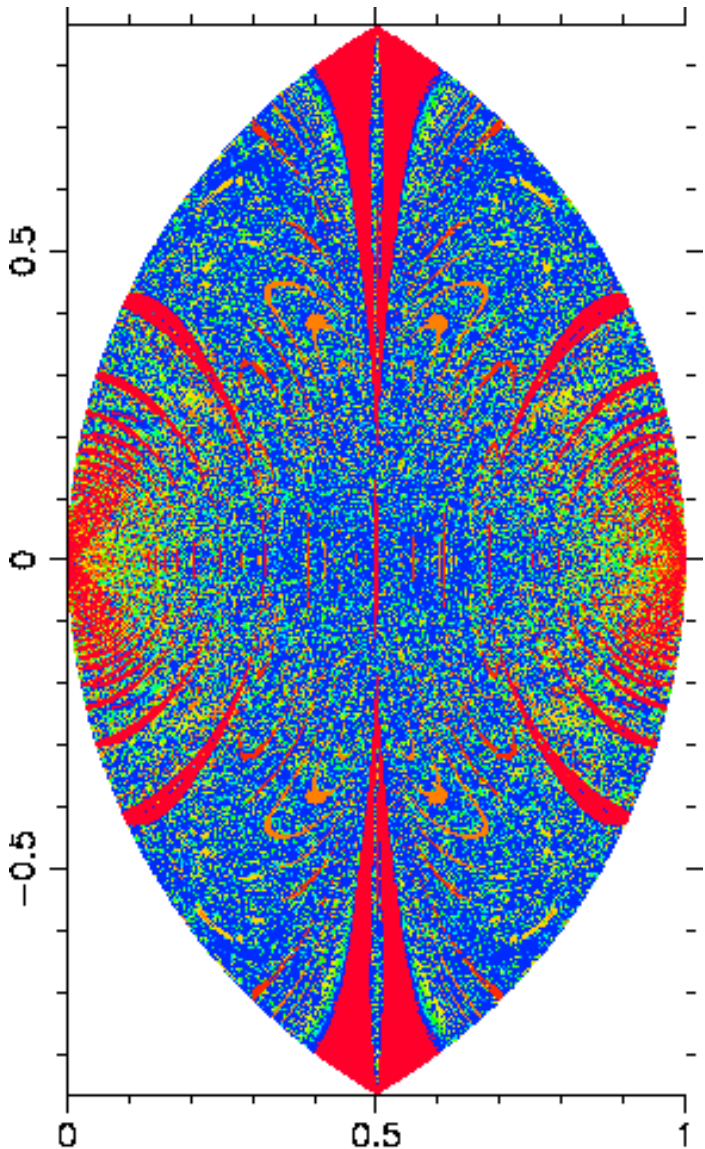}
 \caption{The life time of gravitational three body systems containing all 
possible of orbits that were initially at rest.
The initial configuration of the three body system is such that 
one star is located at $(0,0)$, the second one at $(1,0)$, and
the third one at a coordinate $(x,y)$. 
The colour pixel at $(x,y)$ shows then the time elapsed from the initial 
configuration to the disintegration of the system under Newtonian gravitation.
Red denotes short lived systems, decaying within one crossing time,
orange systems decaying before two crossing times, etc.
The dark blue colours represent systems with the 
longest lifespans (over 13 initial crossing times). 
The pixel size is $1.25\cdot 10^{-3}.$ The image of the complete
shield is symmetric, because we start with all 
three objects at rest. Note the wide resonant bands causing a rapid
disintegration of the system. Also note how narrower bands abound
between wide bands.}
\label{FigShield}
\end{figure*}

The two body problem in a Newtonian gravitational potential is simple 
enough to be solvable in a closed form. Adding a third body to the system 
complicates matters by producing, in general, non-analytic solutions.
Over a century ago \citep{poi}
realized the chaotic nature of the few body system, but only with 
the advance of computer power and more sophisticated algorithms has it been 
possible to give more quantitative statements in the three body problem 
\citep{heg,mik94,aar}.

Here we introduce a simple way to illustrate and study the fractal nature of 
the general three body system. The method we describe provides an important 
tool for visual and qualitative studies of the three body problem. 
It allows us to handle and analyze {\it all\/} possible three body configurations 
within our computational accuracy. The results visualised in this manner     
do not suffer of projection effects.
With this method it is also easy to extend our study to include 
the influence of different initial conditions such as different velocities 
or masses. These extentions will help us understand the chaotic nature of 
the three body systems both in much greater detail and in a more 
comprehensive way. We also discuss shortly the numerical sensitivity of
general three-body problem and show that this may indicate the presence of the Arrow
of Time within these systems.

\section{Decay time-scale and fractality} 
\begin{figure*}
   \includegraphics[width=1.0\columnwidth]{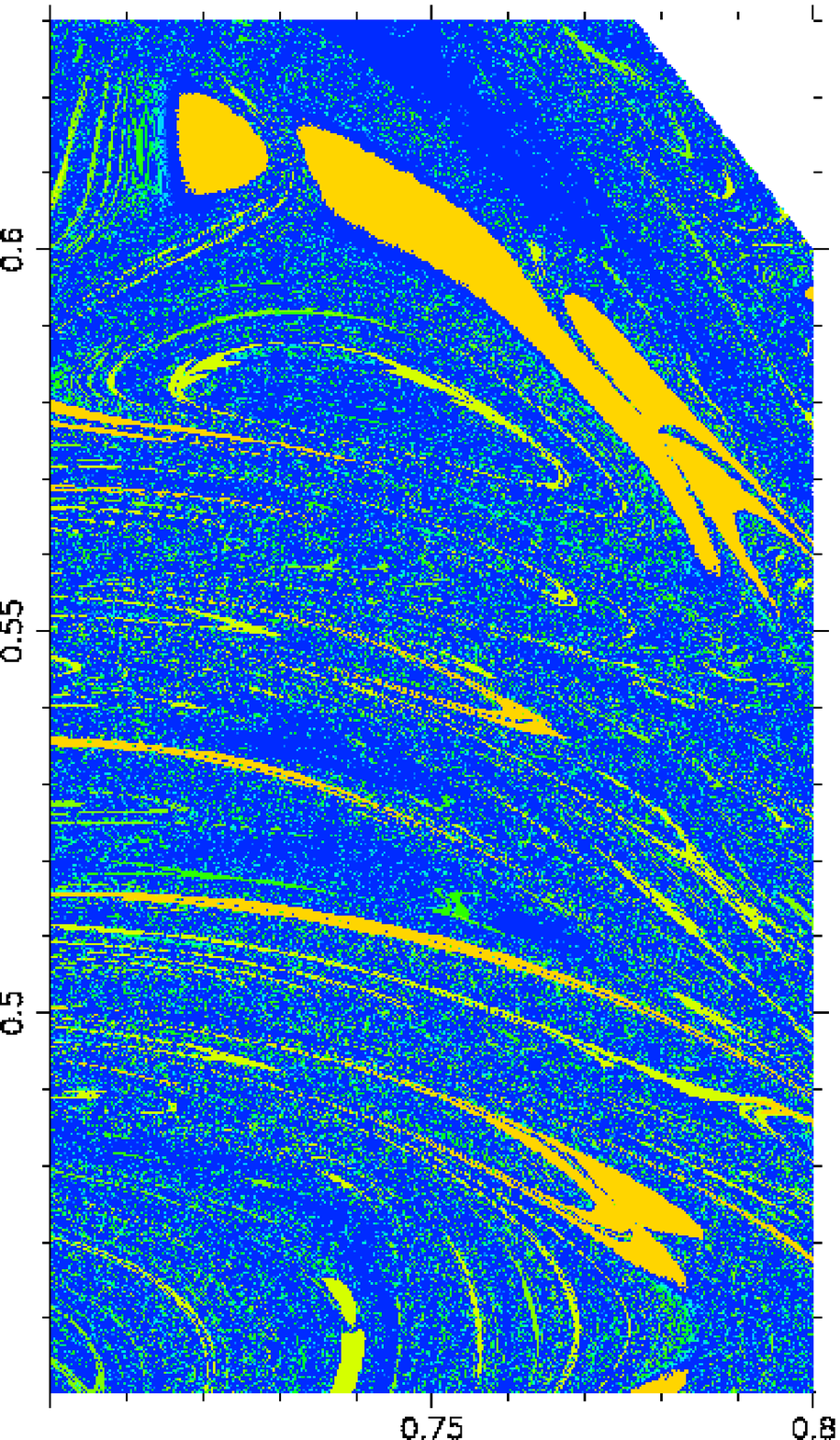}
 \caption{The life time of gravitational three body systems, a zoomed in image 
of Fig. \ref{FigShield}. The pixel size here is $2.5\cdot 10^{-4}$.
Colours as in Fig. \ref{FigShield}.
Note how this region appears quite 'stable' in Fig. \ref{FigShield},
yet in this
image it has broken into hundreds of narrow bands of instability.
Note that the large yellow area breaks into three major parts. This is a
universal feature for the loops. They correspond to the different object
being ejected from the system. The point where these three parts meet is 
a triple collision point. The boundary of the yellow area is sharp 
from inside. If one approaches the boundary from the outside one encounters
a number of orbits growing in density and stability as one approaches the
sharp boundary.}
   \label{FigZoom1}  
\end{figure*}

We have considered the simplest initial state in an equal mass three-body 
system under Newtonian gravitation. The calculations start at rest 
and have no angular momentum. We have used in our calculations
a fully chain-regularized Bulirsch-Stoer integrator 
\citep{mik90,mik93,mik96}.

\begin{figure*}
   \includegraphics[width=1.75\columnwidth]{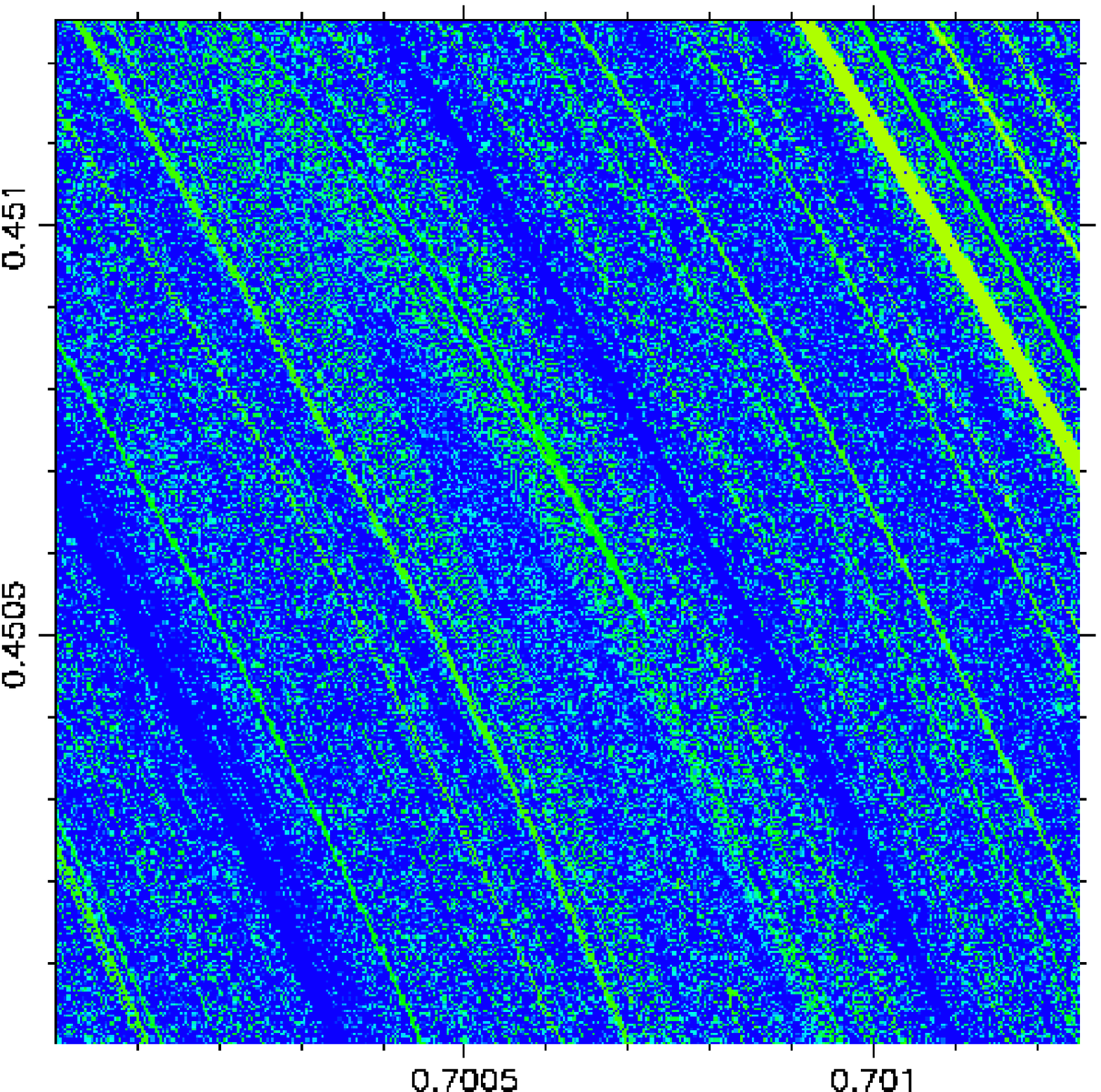}
   \caption{The life time of gravitational three body systems, a zoomed image 
of Fig. \ref{FigZoom1} showing the smaller details. The pixel size here is 
$3.125\cdot 10^{-6}$.
The colour scale is linear and is extended to 20 crossing times. The dark 
blue colours show orbits that have not disintegrated in 20 crossing times. 
The light green band in the right upper corner of this image corresponds 
to a disintegration in about 7 crossing times.}
   \label{FigZoom2}
\end{figure*}

To visualize our simulations we use an extended version of an homology
map \citep{age}. In this 
coordinate system one body resides initially at $(0,0)$, the second one 
is at $(1,0)$ and the third one is at $(x,y)$.  A shield shaped
area contains all possible values of $(x,y)$, spanning 
in abscissa from 0.0 to 1.0 and in ordinate from $-\sqrt{3}/2$ to
$\sqrt{3}/2\approx 0.866.$ A three body configuration can thus be 
unambiguously specified by a single point in this map. 
The unit time-scale of our ($G=1$) system is 
$t={1\over 2\pi}{\left(R/ 1\rm AU\right)}^{3/2} \left({M/ M_{\sun}}\right)^{-1/2}\,\rm years,$
where $R$ is the longest separation of the initial triangle and $M$ is the 
mass of the objects. 

Except for a few double or triple collision points \citep{ume}, all
our three-body systems will eventually break up.
The escaping body will have a non-negative total orbital energy. To define
the breakup time of the system we also require that the escaping body 
must be at a distance of $r\ge 1$ from both remaining bodies. 
This is necessary because during strong interactions the total energy of a 
body may, as expected, become momentarily highly positive.

Because the escaping body often has only a mildly
positive energy and because the quantity we measure is time (the conjugate 
coordinate of energy in phase space), our measurement is in some respect
analogous to a phase space cross-section of a triple system. 

The lifetime of the system may now be plotted as a function of the initial
configuration.
A complicated structure emerges (Fig. \ref{FigShield}). As a curiosity, the overall
pattern shows a nice resemblance with a traditional African Maasai warrior shield
\citep{hun}. Since we consider an equal mass system, there is a left-right 
symmetry in the diagram. The non-zero initial velocity, eg. for the third body
will in general break the up-down symmetry.

\subsection{Resonances} 
On the largest scales the following
regions of instability can be identified. A configuration initially close
to an isosceles triangle (1:1 resonance) is generally highly unstable 
(the vertical red band close to the vertical symmetry line of the map). 
Systems close to odd-digit
orbital resonances  (1:3, 1:5, 1:7 etc) are highly unstable, because a very 
close approach takes place at the first close triple encounter. These are the
bands growing denser towards the left and the right corners of the map. 
Note also the branching of the 1:7 and higher order resonances. 
Furthermore, resonances 1:2, 1:4 etc. are completely missing, because
the phase in these cases is not favorable at the first close encounter
to a rapid break-up of the system \citep{sas}.
Systems breaking up due to the second close interaction show up as 
long orange arcs. The family of these arcs starts from the central part of the map and 
extends towards the edges of the map.  A second system of arcs
between the previous set and the 1:3 resonance
is also due to the odd resonances at the second encounter. 
A band due to similar resonances (nearly vertical bands) can be seen at 
the edge of the 1:1 resonance.

All the major resonances appear to 
avoid a sea of apparent tranquility around the surroundings of 
$(x,y)=(0.7, 0.5)$ and the corresponding points in the other quarters 
of the shield. 

\subsection {Local structure} 
Locally the boundaries of these resonance regions are sharp from the 
inside. Approaching the boundary from outside one encounters ever denser 
loops and stripes. This structure appears selfsimilar and locally fractal.

Areas of higher stability are also present in Fig \ref{FigShield}. These fall into two 
categories. The first set are isolated deep trenches on large scales at 
about $(0.70, 0.64)$ and $(0.65, 0.15)$ and at respective symmetric 
points of the shield. The second set better visible on small scales
are configurations just outside some instability regions e.g. around the 
upper branch of the 1:7 resonance at (0.90, 0.23) or below the yellow eye at 
$(0.75, 0.60)$ and the 
corresponding symmetric points.

Figures \ref{FigZoom1} and \ref{FigZoom2} show progressive levels of zooming into these areas.
Figure \ref{FigZoom2} represents a 5 by 5 pixel area in the lower left corner of
Figure \ref{FigZoom1}. The width of the strips seen in 
Figure \ref{FigZoom2} is comparable to the resolution 
element or $3.125\cdot 10^{-6}$. These stripes are more easily seen if the
image is viewed at a small angle. It is clear that a tiny change in 
the initial conditions may cause a substantial difference in the
lifetime and in the evolution of the system. Further details of the orbital
behavior within these areas is beyond the scope of this paper, except
that we may state that these configurations appear to be more stable.

\begin{figure}
   \includegraphics[width=1.0\columnwidth]{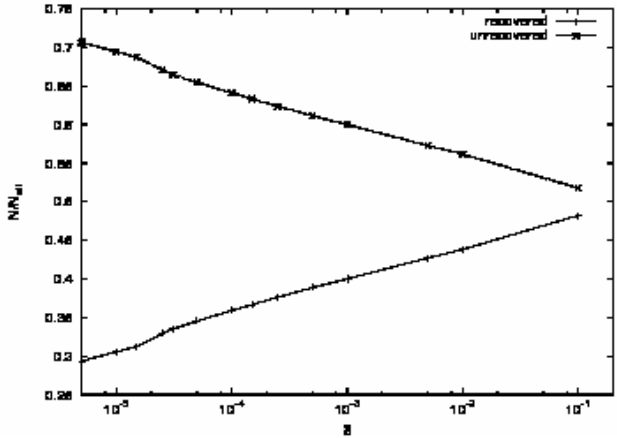}
   \caption{The number fraction of recovered and unrecovered
orbits as a function of required accuracy $\delta$. At
around $\delta = 10^{-5}$ one come to the limit of decimals in the result
 files.}
   \label{deltafig}
\end{figure}

\subsection {Global fractality} 
To investigate the ratio of stable (locally homogeneous) regions to
the chaotic regions we compared, $t_{\rm o}$, the ejection time of an orbit
to $t_{\rm med}$, the median of 8 nearby orbits located on a rectangular grid
with total width of $3.75\cdot 10^{-3}$. The eight configurations are 
separated by $\epsilon=\pm 1.25\cdot 10^{-3}$ in each of the $x$ and $y$ 
coordinates from the initial point. 

If $|t_{\rm med}-t_{\rm o}|\le 0.5,$ we considered the point to have an locally
homogeneous neighbourhood. Otherwise it was considered to have high sensitivity
to initial conditions, a characteristic of chaos. Of the 21\,699 cases 
compared in this manner 6212, or $p_\epsilon=28.6$ per cent were in locally
homogeneous.
The remaining 71.4 per cent of positions turned out to be highly sensitive to initial 
conditions at this resolution implying that on the scale of 
$\epsilon=1.25\cdot 10^{-3} $ the fractal area covers about 
71.4 per cent of all the systems.
The implication of this is that if one is able to determine the initial state
of a triple system to an accuracy of $\epsilon$, one will, with the stated 
probability, be unable to tell the time-scale of the ejection of the third body 
to within half a crossing time.

\begin{figure}
   \includegraphics[width=1.0\columnwidth,angle=0]{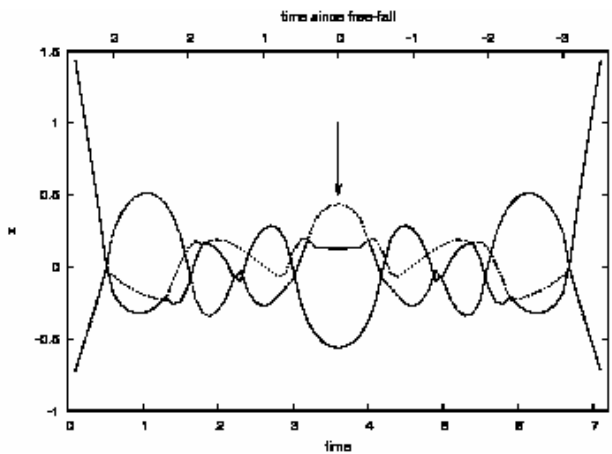}
   \caption{An example of a time symmetric orbit. The
 orbit is calculated first from free-fall position to escape (part not shown here) and then
correctly back in time to the original free-fall situation which is found at
$t=3.6$ and indicated by an arrow. Finally the integration was
continued until the system breaks up again at $t=7.2$.}
   \label{FigSym}
\end{figure}
\begin{figure}
    \includegraphics[width=1.0\columnwidth,angle=0]{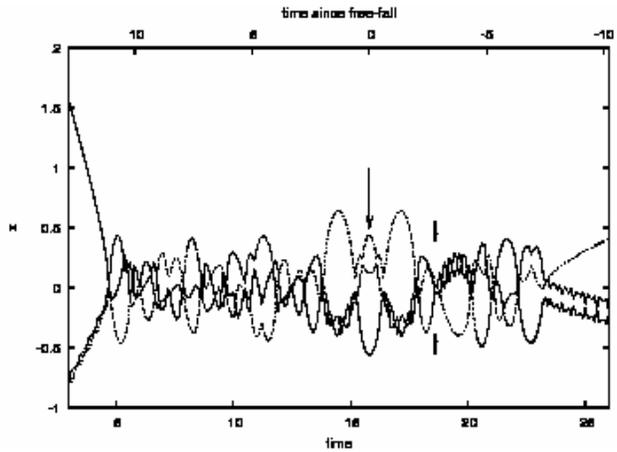}
    \caption{An example of an orbit which is asymmetric with
respect to time. This orbit is also traced back to its free-fall moment
 at $t=15.8$ (marked with long arrow) but the motion after this differs
from the previous history at $t=18.6$ (marked with two arrows) leading
in this case dramatically different break up.}
   \label{FigAsym}            
\end{figure}

If our image is fractal one should find more and more (local) order as one 
decreases the resolution element. This will be difficult to do due to 
restrictions in computation time. We can increase the resolution and 
see if the amount of local homogeneity decreases in a self similar way.
For bin sizes 1, 2, 3, 4, 5 and 10 times the original one we find a
decrease in local homogeneity, $p_\epsilon,$ 
28.6, 16.8, 14.9, 13.8, 13.5 and 8.6 per cent, respectively.
To within measurement errors $p_\epsilon \propto \epsilon^{-0.5}$.
We find the same exponent if we study the detailed areas shown in Figures 
\ref{FigZoom1} and \ref{FigZoom2}. The Hausdorff dimension \citep{hau} of our image is thus
$$D_{\rm H}=1.5.$$

\begin{figure*}
   \includegraphics[width=2.0\columnwidth,angle=0]{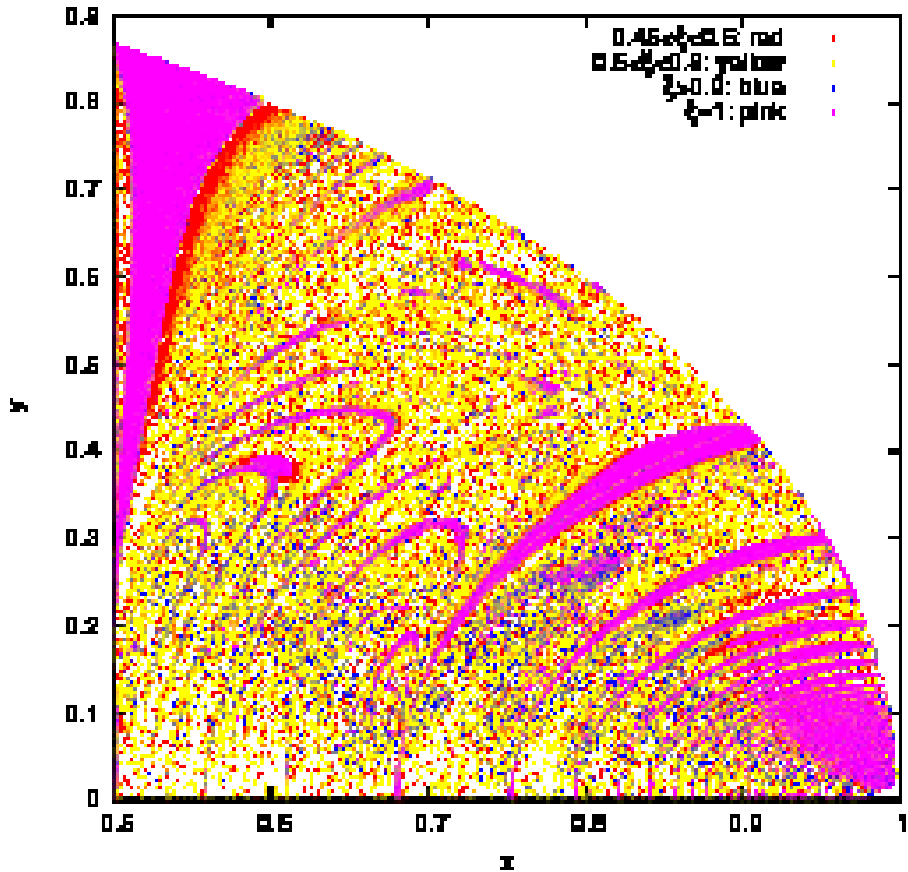}
   \caption{The homology map for most recovered orbits. The structure
of the map is very similar to the one obtained by ejection times in
Fig. \ref{FigShield}. Here the colour
indicates $\xi$, the amount of time symmetry, for each orbit. Notice how
this value tend to increase when initial configuration depart from the
resonance bands.}
   \label{FigMapSymAsym}
\end{figure*}

\section {Time reversal and the Arrow of Time} 
The initial conditions we choose for our systems have the important 
mathematical property that the orbits we calculate are time symmetric 
with respect to the origin.  This implies that the triple systems we 
calculate are transient in the sense that the body expelled 
must have been once captured by the very same bodies forming the final 
binary system (and even in the very same phase but with an opposite sign!). 
Physically this is not quite true, because even a small disturbance to the 
system would cause the system to loose this property. This also implies that 
a simple time reversal at the time of escape will not be a useful indicator 
of the errors involved in these calculations.

It is possible that the initial conditions are not recovered by
two reasons: the calculations may not be accurate enough, and
errors accumulate to the extent where the initial conditions are missed on the way back.
The second possibility is that the calculation is accurate enough, but
the individual system is so sensitive that it is practically impossible
to find the initial conditions again. See \citet{aar} for a
discussion of predictable and non-predictable orbits.

To ensure independence from the method our calculations were done with
two different 
methods. The method uses a code consisting of a Bulirsch-Stoer      
integrator with the chain regularization algorithm \citep{mik93}. The
second code uses a simple leapfrog integrator to logarithmic   
Hamiltonian \citep{mik99}. Results obtained are in
agreement but in analysis we prefer the latter code because of its
constant stepsize.

The orbits were calculated in few stages. First a forward 
integration was carried out until one of the bodies escapes. Here one
should notice that criterion for recognizing an escape is not trivial.
An escape is suspected if the ratio between the
semi-major axis of one body orbit and the semi-major axis of
the binary (i.e. two bodies close to each other) is large enough. We
have selected this ratio to be 100. In addition to this requirement
the system has to fulfil one of the two following conditions. An
escape is identified if the third body is on an hyperbolic orbit
\emph{and} is moving away from the binary. Alternatively, if one of the bodies
gets so far that the ratio exceeds $\ge 2000$ we consider this an escape,
even though a proper hyperbolic orbit is not present. These escape
criteria follow \citet{ano}, which includes also a list of
different possible escape criteria. We have also tried smaller values
for the escape criteria with no apparent effect on our results.

At the second stage the escape 
velocities were reversed and a reverse orbit was calculated until the
system reached its initial conditions again. After that the simulation
has carried on until the system breaks up again.

The definition of when the initial conditions has been recovered
depends on the accuracy of    
the computer and software. We find that by using double precision we can    
expect results to be correct at least with accuracy $\sim 10^{-7}$. To
avoid rounding errors we consider only five significant digits. Therefore  
we require that the separation between original initial state and its
recovered counterpart must be less than $\delta = \pm 5.0 \cdot
10^{-5}$. The choice of this limit
affects naturally the ratio of the recovered and unrecovered
orbits. The ratio is shown in Figure \ref{deltafig} where different values of $\delta$ has been       
applied to the dataset.

\begin{figure}
   \includegraphics[width=1.0\columnwidth,angle=0]{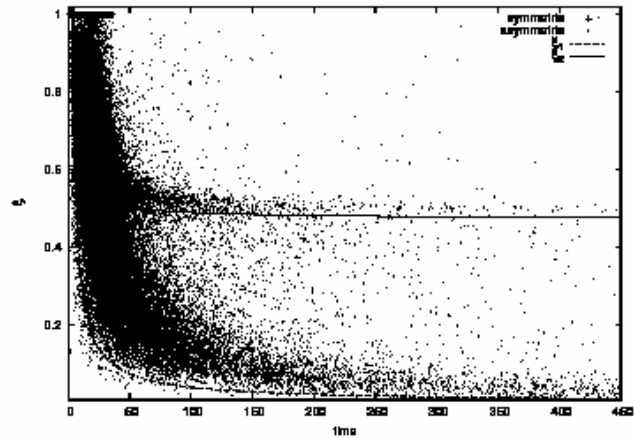}
   \caption{The symmetricity, $\xi$, of the orbit as a function of the
lifetime. The functional forms $\xi_1(t_{\rm tot})$ and $\xi_2(t_{\rm tot})$ for two distinctive populations of
time asymmetric orbits are also shown (see text). Totally time symmetric orbits
form short horizontal line at $\xi = 1$.}
   \label{FigTbs}
\end{figure}

If one would have an infinite accuracy one would obtain an orbit where
the motion is exactly the same before and after the
free-fall moment -- in the other words the orbit is time symmetric.
This is a fact which is based on the equations of motion.
However, in most cases in our numerical calculations this is not the case even though the
free-fall moment can be recovered: most orbits actually
forget their past after the free-fall moment producing an asymmetrical
time evolution. This difference is illustrated with examples in Figure
\ref{FigSym} for time symmetric and Fig. \ref{FigAsym} for time asymmetric orbits. We define a new quantity
$\xi$ as time ratio of the a recovered part to the orbit's time of
disruption $t_{\rm tot}$. If $\xi=1$ the orbit is exactly the same in
the past and in future from the free-fall    
moment. In Fig. \ref{FigMapSymAsym} the value of $\xi$ is presented for each   
orbit identified by its initial position on the homology map. The
orbits marked by pink colour represent the totally symmetrical evolution while other
colours indicate binned values of $\xi$ for asymmetric orbits. In red
pixels the length of the symmetric
part is between $0.45-0.60$ of $t_{\rm tot}$. In the yellow pixels symmetry
covers $0.6-0.9$ of $t_{\rm tot}$ and  
at the blue pixels the orbit is over 90 but less than 100 per cent
symmetric. The white areas -- distributed quite equally
everywhere on the map outside the resonances -- loose their symmetry    
faster than within $0.45$ of $t_{\rm tot}$ or they represent cases where the
orbit has not been recovered. The bins for $\xi$ in Fig.
\ref{FigMapSymAsym} are based on the distribution of orbits in Figure
\ref{FigTbs} which we will discuss later.

According to Fig. \ref{FigMapSymAsym} one will find the exactly time
symmetric behaviour within resonance areas where orbits have
relatively short lifetimes. But is there any relation between the
lifetime of the asymmetric orbits and
their sensitivity to the numerical errors? This question is answered
in Figure \ref{FigTbs} where we present the amount of 
symmetry, $\xi$, as a function of disruption time of the system. It appears that the
population of asymmetric orbits are divided into two, partially
overlapping, groups. Both of these groups have distinctive functional
form $\xi_i(t_{\rm tot}) = a_i t_{\rm tot}^{-n_i}+b_i$ and these sets appear to have quite
sharp edges on the minimum side. Population I ($\xi \ga 0.45$) seems to be direct continuum of time symmetric
orbits ($\xi = 1$) while the lower edge of the curve
has approximately parameters $a_1=0.5, n_1=0.7$ and $b_1=0.47$. These orbits appear to be located
just outside the resonance areas on the homology map and the value of $\xi$
increases when the initial configuration moves farther from the
resonance bands.

\begin{figure}
   \includegraphics[width=1.2\columnwidth,angle=0]{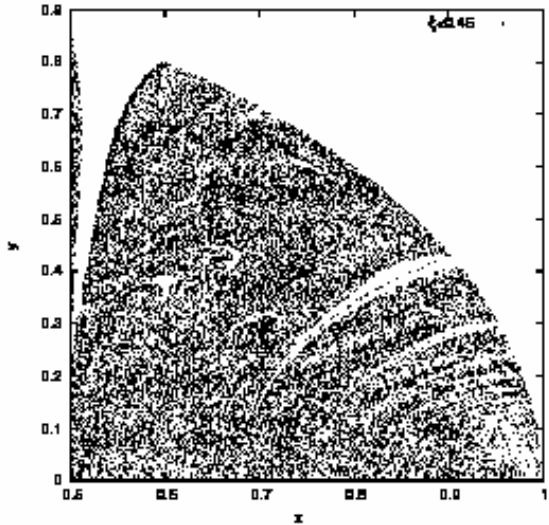}
   \caption{The homology map for recovered orbits with $\xi \le 0.45$.
These orbits do not form any grouping as seen for higher values of
$\xi$ in Fig. \ref{FigMapSymAsym}.}
   \label{FigMap045}
\end{figure}

Most orbits however belong to population II ($0<\xi<1$) which is quite equally
distributed over the homology map, clearly further from the resonance bands. The
lower limit of $\xi$ in this population obeys approximate relation
$a_2=2.5, n_2=0.9$ and $b_2=0$. On the homology map
(Fig.\ref{FigMap045}) this population is strongly intermixed
with non-recovered orbits but the latter ones do not show any
clear functional form for $\xi = \xi(t_{\rm tot})$.

We have found out that even in quite a simple dynamical system such the
classical equal-mass free-fall three-body problem, actually the Arrow
of Time shows up. Considering the analytical point of      
view the autonomous form (i.e. no explicit time dependency) 
of the equations of motion and the lack of dissipation suggest that
this arrow would not be there if we could solve the problem exactly.    
However, in nature totally unperturbed systems do not exist and even a
minute perturbation, just like a small computing error, is adequate to 
define a definite sense of time.

How do we find out what is the forward sense of time in the three-body 
problem? The answer is provided by the concept of entropy again. Not  
the classical Boltzmann entropy, but Kolmogorov-Sinai entropy
\citep{kol,sin}. This    
concept of entropy applies to small systems. It is related to the    
change of the volume of phase space occupied by systems which are    
initially very close to each other. It measures the loss of          
information about the initial conditions or the growth of disorder as
time goes on. We have previously shown that Kolmogorov-Sinai entropy  
increases with time in the three-body problem, and that it thus tells
us what is future, and what is past \citep{hei}. The  
Kolmogorov-Sinai entropy increases towards future time. 

There are other examples of determining the sense of time in the       
three-body problem. In three-body scattering a single body collides    
with a binary, and after some orbital evolution, one of the bodies   
escapes and leaves again a binary. Let us say that the initial binary
has zero eccentricity. The final binary has an eccentricity $e$ that  
follows the distibution $f(e)=2e$            
\citep{jea,heg}. Thus the probability   
that the final binary escapes with zero eccentricity is zero. In a   
diagram depicting such a scattering event \citep{hut}  
it is immediately clear which is the direction of orbital motion even
if it is not indicated. Although some of our orbits perform time    
symmetric past and future they resemble a minority in our ensemble and
it is somewhat reasonable to consider the Arrow of Time in the universe as a 
statistical quantity. One may also ask if these time symmetric orbits 
really appear at all in true nature where small perturbations are    
always present.

The microscopic world consists of many small subsystems similar to the 
classical three-body system. It is our conjecture that our conclusions 
would apply also to such systems, and that Kolmogorov-Sinai entropy    
would actually provide the Arrow of Time for the microworld.

There have previously been a number of papers cosidering a relation between the
Kolmogorov-Sinai entropy and the Boltzmann thermodynamic entropy.
These studies, which have been done in other fields of physics. They have
either tried to address the relation between these
two quantities, albeit such relation still remains unknown,
\citep{lat,bar} or attempted to find a direct equivalence between the
dynamical instability and the 
thermodynamic irreversibility \citep{els,yuk,rui}. If these 
relations indeed exist, they endorse our conclusion.

\section {Noise estimates} 
Since noise estimates are difficult to perform in these kinds of systems.
We followed the simpler practice of reducing the integration step 
by a factor of 2. We noticed only a slight sharpening in the boundaries, but no
qualitative differences nor any unexpected quantitative differences.

A second strong test we have performed is for the robustness of the method.
We ran a na\"\i ve simulation with non-regularized equations of motion but 
with a short variable steps of integration. The code is very 
different in nature from our main code. Yet the results are very similar
showing the main resonant structures. 
On small scales similar fractal-like patterns appear as with our main code.

\section {Discussion} 
The results from \citet{hei} and \citet{ano}, suggested that
a homology triangle-like representation is an ideal means to describe
the evolution or the final solutions of a triple system, and that it would 
show structure both due to global and local sensitivity to initial conditions.
The geometry of escape times which we have calculated shows rich structure 
featuring resonance bands at various levels and areas of apparent higher 
stability as well as areas with clearly a fractal structure. 

The time-scales in question for solar mass objects and an intial length scale of
$R=1\,\rm pc$ correspond to $t \sim 1.6\cdot 10^7 \,\rm{years}.$
The implication is that triple stellar systems forming initially
from distant stars may still be in a multiple system after a few tens of
million years (the blue areas in Figure \ref{FigShield} imply time-scales 
$>1.6 \cdot 10^8 \,\rm{years}$).
On the other hand one would expect stars being ejected by this method
from young systems to form loose associations at the latest in about 
a ten million years. If the stars were closer ($R=0.1\,\rm pc$) then the
breakup times would be a factor 30 shorter. On the other hand if the stars 
were not initially at rest then these time-scales are likely to be somewhat 
longer based on previous work (see e.g. \citet{sas,val}). 

Galaxies at distances of $\sim 1\,\rm Mpc$, and with masses 
$M\sim 50\cdot 10^9 \rm M_{\sun}$ have a typical time-scale of $10^{10.5}$ years, 
so these are expected to be still these initial orbits, even if they
were near the resonance configuration.

\section*{Acknowledgments}
This work has been funded by Finnish Cultural Foundation and Jenny and
Antti Wihuri Foundation.


\begin{thebibliography}{99}

\bibitem[\protect\citeauthoryear{Aarseth et al.}{1994}]{aar} Aarseth S.J., Anosova J.P., Orlov V.V., Szebehely V.G., 1994, Celest. Mech., 58, 1

\bibitem[\protect\citeauthoryear{Anosova}{1986}]{ano} Anosova J.P., 1986, Ap\&SS, 124, 217

\bibitem[\protect\citeauthoryear{Agekian \& Anosova}{1967}]{age} Agekian T.A., Anosova J.P., 1967, SvA, 44, 1261

\bibitem[\protect\citeauthoryear{Baranger et al.}{2001}]{bar} Baranger M., Latora V., Rapisarda A., 2001, Chaos Solitons and Fractals, 13, 471

\bibitem[\protect\citeauthoryear{Elskens \& Prigogine}{1986}]{els} Elskens Y., Prigogine I., 1986, PNAS, 83, 5756

\bibitem[\protect\citeauthoryear{Hausdorff}{1919}]{hau} Hausdorff F., 1919, Mathematische Annalen 79(1--2), 157

\bibitem[\protect\citeauthoryear{Hein\"am\"aki et al.}{1999}]{hei} Hein\"am\"aki P., Lehto H., Valtonen M., Chernin A.D., 1999, MNRAS, 310, 811

\bibitem[\protect\citeauthoryear{Heggie}{1975}]{heg} Heggie D.C., 1975, MNRAS, 173, 729

\bibitem[\protect\citeauthoryear{see e.g. Huntingford}{1961}]{hun} Huntingford G.W.B., 1961, The Journal of the Royal Anthropological Institute of Great Britain and Ireland, Vol. 91, No. 2., 251

\bibitem[\protect\citeauthoryear{see e.g. Hut \& Bahcall}{1983}]{hut} Hut P., Bahcall J., 1983, ApJ, 268, 319

\bibitem[\protect\citeauthoryear{Jeans}{1919}]{jea} Jeans J., 1919, MNRAS, 79, 408

\bibitem[\protect\citeauthoryear{Kolmogorov}{1958}]{kol} Kolmogorov A.N., 1958, Dokl. Acad. Nauk. SSSR, 119, 861

\bibitem[\protect\citeauthoryear{see e.g. Latora et al.}{1999}]{lat} Latora V., Baranger M., 1999, Phys. Rev. Lett. 82, 520

\bibitem[\protect\citeauthoryear{Mikkola}{1994}]{mik94} Mikkola S., 1994, MNRAS, 269, 127

\bibitem[\protect\citeauthoryear{Mikkola \& Aarseth}{1990}]{mik90} Mikkola S., Aarseth S.J., 1990, Celest. Mech, 47, 375

\bibitem[\protect\citeauthoryear{Mikkola \& Aarseth}{1993}]{mik93} Mikkola S., Aarseth S.J., 1993, Celest. Mech, 57, 439  

\bibitem[\protect\citeauthoryear{Mikkola \& Aarseth}{1996}]{mik96} Mikkola S., Aarseth S.J., 1996, Celest. Mech, 64, 197
 
\bibitem[\protect\citeauthoryear{Mikkola \& Tanikawa}{1999}]{mik99} Mikkola, S., Tanikawa K., 1999, MNRAS, 310, 745

\bibitem[\protect\citeauthoryear{Poincar\'e}{1892}]{poi} Poincar\'e H., 1892, Les M\'ethodes Nouvelles de la M\'echanique Celeste. Gauthier-Villars, Paris 

\bibitem[\protect\citeauthoryear{Prigogine \& Stengers}{1984}]{pri} Prigogine I., Stengers I., 1984, Order out of Chaos. Bantam books, New York

\bibitem[\protect\citeauthoryear{Ruiz \& Tsallis}{2007}]{rui} Ruiz G., Tsallis C., 2007, Physica A, 386, 720

\bibitem[\protect\citeauthoryear{Saslaw \& Valtonen}{1974}]{sas} Saslaw W.C., Valtonen M.J., Aarseth S.J. 1974, ApJ, 190, 253 

\bibitem[\protect\citeauthoryear{Sinai}{1959}]{sin} Sinai Ya.G., 1959, Dokl. Acad. Nauk. SSSR, 125, 1200

\bibitem[\protect\citeauthoryear{Umehara \& Tanikawa}{2000}]{ume} Umehara H., Tanikawa K., 2000, Celest. Mech, 76, 187

\bibitem[\protect\citeauthoryear{Valtonen \& Karttunen}{2006}]{val} Valtonen M.J., Karttunen H., 2006, The Three-Body Problem. Cambridge Univ. Press, Cambridge

\bibitem[\protect\citeauthoryear{Yukalov}{2003}]{yuk} Yukalov V.I., 2003, Physica A, 320, 149

\end{thebibliography}
\end{document}